\documentclass[pra,aps,twocolumn]{revtex4}
\usepackage{graphicx}
\usepackage{dcolumn}
\usepackage{booktabs}
\usepackage{amsmath}

\usepackage{bm}

\makeatletter
\renewcommand{\@seccntformat}[1]{} 
\renewcommand{\section}{%
  \@startsection{section}{1}{\z@}%
  {0.5\baselineskip}%
  {0.1\baselineskip}%
  {\normalfont\bfseries\noindent}%
}
\def\@sect#1#2#3#4#5#6[#7]#8{%
  \ifnum #2>\c@secnumdepth
    \let\@svsec\@empty
  \else
    \refstepcounter{#1}%
    \protected@edef\@svsec{} 
  \fi
  \@tempskipa #5\relax
  \ifdim \@tempskipa>\z@
    \begingroup
      #6{\@hangfrom{\hskip #3}
        \interlinepenalty \@M \bfseries #8\@@par}%
    \endgroup
  \else
    \def\@svsechd{#6{\hskip #3\relax\@svsec #8}}%
  \fi
  \@xsect{#5}}
\makeatother

\begin{document}


\title{Unidirectional and collective emission of integrated quantum emitters}

\author{Jun Ren$^{1,2,3}$}
\email{renjun@hebtu.edu.cn}
\author{Meng-Jia Chu$^{1,3}$}
\author{Z. D. Wang$^{1,2}$}

\affiliation{$^1$ HK Institute of Quantum Science $\&$ Technology, The University of Hong Kong, Pokfulam Road, Hong Kong, China\\
$^2$ Hong Kong Branch for Quantum Science Center of Guangdong-Hong Kong-Macau Great Bay Area, Shenzhen, China\\
$^3$ College of Physics and Hebei Key Laboratory of Photophysics Research and Application, Hebei Normal University, Shijiazhuang, Hebei 050024, China\\
}

\collaboration{CLEO Collaboration}

\date{\today}

\begin{abstract}
Unidirectional emission holds significant potential for advancing integrated photonics and quantum information technologies. However, the inherent randomness of spontaneous emission fundamentally makes its efficient realization rather challenging. To address this, here we develop a quantitative metric---iso-frequency contour straightness and implement Fourier-transform analysis of radiation patterns to systematically evaluate directional quality of emission in photonic crystal (PhC) slabs. Through structural optimization, we demonstrate single-emitter radiation efficiency enhancement while maintaining low-loss unidirectional propagation. Furthermore, by strategically positioning multi-emitter arrays within PhC platforms, we simultaneously achieve scalable intensity amplification and superradiant emission via cooperative effects. This synergy of photonic band engineering and collective emitter coupling is able to realize unprecedented spatiotemporal coherence control in quantum emitter arrays.

\end{abstract}

\maketitle


\section{Introduction}
Spontaneous emission, as a fundamental process in quantum optics, typically results in isotropic photon radiation due to vacuum fluctuations \cite{Milonni76pr,Scully97book}. This inherent directionality randomness presents both a fundamental challenge and a practical limitation, as controlling emission direction has been a long-standing research objective in light-matter interaction studies \cite{Lodahl04nature,Coenen14NC,Mitsch14NC}. The radiation isotropy can be broken through various approaches, including near-field coupling with adjacent emitters or careful engineering of emitter arrays, leading to anisotropic or even unidirectional radiation patterns \cite{Taminiau08NP,Curto10science}. Such directional emission control carries profound significance beyond fundamental science \cite{Bhatti18prl}, enabling quite promising applications ranging from quantum networking---where directional photon communication between remote nodes is essential \cite{Kimble08nature,Reisere15rmp,Kannan23np}---to advanced laser systems requiring precisely engineered emission patterns \cite{Song10prl,Jiang12am,Schermer15apl}. However, achieving on-demand, high-efficiency directional emission remains rather challenging, particularly in scalable quantum photonic platforms.

Nanophotonic devices have emerged as pivotal tools for manipulating light-matter interactions across quantum optics and integrated photonics \cite{Novotny12book,
Koenderink15science,Rivera20nrp}. Advanced architectures---including hybrid cavity-antenna systems \cite{Curto10science,Kosako10np,Coenen11nl,Coenen14NC,Stella19ap} and atomic arrays \cite{Shahmoon17prl,Garcia17prx,Masson20prl,Fernández22prl} further enable precise modulation of emission directionality. Among these platforms, photonic crystal (PhC) slabs have garnered significant attention for their ability to tailor emitter radiation patterns through guided modes and band structure engineering \cite{Lodahl04nature,Englund05prl,Fujita05science,Arcari14prl}. Recent advances in PhC slab design have further demonstrated their exceptional capability to control directional emission through guided modes, with a particularly promising approach utilizing two-dimensional PhCs featuring saddle-point band structures and linear isofrequency contours to generate highly directional emission channels \cite{Barón24np,Yu19pnas}. Collectively, these low-loss periodic structures establish a robust framework for controlling radiation directionality, bridging fundamental light-matter coupling studies with scalable photonic device applications.

\begin{figure*}
	\includegraphics[width=0.8\textwidth]{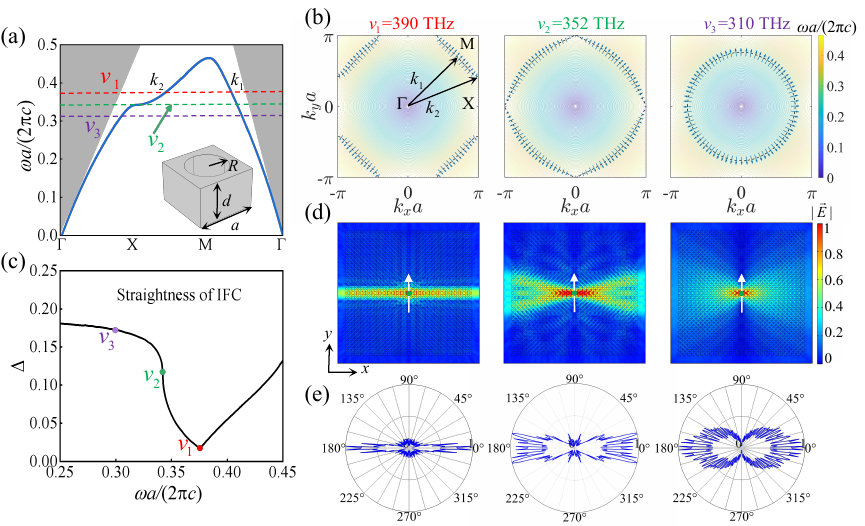}
	\caption{(Color online) \textbf{Directional emission in a PhC slab.} (a) For the first band of the TE-like mode in a square lattice PhC slab, three dashed lines of different colors represent the unidirectional frequency $\nu_1=390~\rm THz$ (red dashed line), saddle-point frequency $\nu_2=352~\rm THz$ (green dashed line), and $\nu_3=310~\rm THz$ (purple dashed line), respectively, with the unit cell structure of the slab shown in the inset. The lattice constant is $a=290~\rm nm$, the thickness is $d=200~\rm nm$, the hole radius is $R=103~\rm nm$, and the dielectric constant is $\varepsilon_{r}=4$. $k_1$ and $k_2$ are the two modes corresponding to the $\nu_1$ frequency. (b) are the IFC plots corresponding the energy band in (a). The black dashed lines represent the IFC for the specific frequencies $\nu_1$, $\nu_2$, and $\nu_3$, and the blue arrows denote the group velocity. The two black arrows indicate the wave vectors corresponding to $k_1$ and $k_2$. (c) The straightness corresponding to the IFC at different frequencies. (d) are respectively the distribution diagrams of the electric field intensity of the dipole (green dot) on the central cross-section of the slab at different frequencies, in which the white arrow indicates the polarization direction of the dipole. (e) show polar plots of the electric field intensity at a distance of $8\sqrt{2}a$ from the dipole under different frequencies, where the radius of the polar plots represents the magnitude of the field strength.}
\end{figure*}

While previous studies on PhC slab-mediated directional emission primarily focused on controlling single-emitter radiation patterns, practical applications demand scalable emission devices involving multiple interacting emitters where quantum correlations fundamentally reshape the emission process (e.g., superradiance and subradiance) \cite{Fujita05science,Minkov18prl,Barón21acsp,Dyakov23prb,Zhu23oe,Ren24prapp,Qian24pra,Barón24np,Chu25npj}. However, the cooperative mechanisms governing unidirectional radiation in emitter arrays remain unclear, particularly regarding how collective states interact with photonic band structures, making this an intriguing research frontier bridging quantum optics and nanophotonics. Moreover, existing metrics for directional emission quality have relied excessively on the straightness of iso-frequency contours (IFCs). The present research reveals that IFC straightness alone cannot fully assess emission quality, which fails to capture unidirectionality under realistic conditions (such as disorder and sample size) and ignores radiative losses from scattering or leaky modes. These limitations necessitate a comprehensive standard rule---a crucial step toward scalable quantum light sources with high directionality and efficiency.

In this work, we develop a comprehensive assessment scenario of directional radiation for emitters embedded in a PhC slab. We introduce key parameters for quantifying the unidirectionality of radiation—specifically, the straightness of IFCs and Fourier transform analysis of the radiation field. Our results demonstrate that this set of parameters effectively evaluates the quality of unidirectional emission when optimizing structural parameters of the slab. Furthermore, for the collective radiation effects of multiple emitters in PhC slabs, we investigate not only classical spatial interference of fields but also coherent cooperative radiation phenomena, such as superradiance. By precisely controlling the spatial distribution of emitters, we show that the PhC slab platform simultaneously achieves low-loss unidirectional emission, scalable intensity enhancement, and superradiant emission. This highlights the unique synergy between photonic band engineering and multi-emitter coupling, enabling unprecedented control over both spatial and temporal coherence in quantum emitter arrays.

This paper is organized as follows: In the Results section, we introduce metrics for evaluating directional emission quality—specifically, the straightness of iso-frequency contours (IFCs) and Fourier transform analysis of radiation patterns—and outline strategies to enhance single-emitter efficiency. We then explore collective emission effects in multiple emitters. In the Discussion section, we elaborate the impact of symmetry on unidirectional radiation robustness in scalable emitter arrays, followed by an analysis of directional emission in triangular-lattice photonic crystal slabs. The Methods section details the guided-mode approach, Green’s tensor formalism, and Fourier transform techniques underpinning our study.

\section{Results}
\textbf{Directional emission of a single emitter}. Here, we consider a square lattice of air cylinders in a slab with dielectric constant $\varepsilon_{r}=4$. The unit cell structure is shown in the inset of Fig.~1(a), with lattice constant $a=290~\rm nm$, slab thickness $d=200~\rm nm$, and air cylinder radius $R=103~\rm nm$. For this photonic crystal slab, there are two types of eigenmodes, the radiation mode and the guided mode. Usually, the radiation mode exists in the light cone, which radiates electromagnetic waves outside the slab and cannot exist stably in the slab. One exception is BIC, which has a stable bound state in the continuous spectrum due to symmetry \cite{Minkov18prl}. Outside the light cone, theoretically, all modes have infinite quality factors, that is, they can all exist stably in the photonic crystal slab. This work focuses on the latter case, i.e., the guided mode. We can use the Guided Mode Expansion (GME) method to solve its band structure outside the light cone. The details of the GME method can be found in the METHODS section.

The blue line in Fig.~1(a) shows the first energy band for TE-like modes outside the light cone, where the shaded region denotes the light cone. The radiation directivity of a guided mode can be assessed by examining its iso-frequency contour (IFC) and the corresponding group velocity directions \cite{Yu19pnas}. In Fig.~1(b), we highlight (with black dashed lines) the IFCs at three typical frequencies within the first Brillouin zone: $\nu_1=390$~THz, $\nu_2=352$~THz, and $\nu_3=310$~THz, which correspond to the colored dashed lines in Fig.~1(a). Among these, $\nu_1$ is termed the unidirectional frequency due to its nearly straight IFC, while $\nu_2$ represents the X-point frequency, where the band structure exhibits a saddle point---a singularity associated with a divergent local density of states (LDOS) and consequently enhanced spontaneous emission. In contrast, $\nu_3$ serves as a reference case, demonstrating nearly isotropic radiation for comparison. In Fig.~1(a), $k_1$ and $k_2$ are the two modes corresponding to the unidirectional frequency $\nu_1$, with the directions of their wave vectors shown by the black arrows in Fig.~1(b). A more detailed description of these two eigenmodes can be found in the DISCUSSION section.

In Fig.~1(b), the length and direction of the blue arrows represent the magnitude and orientation of the group velocity at each wavevector $\vec{k}$, which can be calculated as
\begin{equation}
    \vec{v}_g = \frac{\partial \omega(\vec{k})}{\partial k_x}\hat{x} + \frac{\partial \omega(\vec{k})}{\partial k_y}\hat{y},
\end{equation}
where $\omega(\vec{k})$ describes the dispersion relationship shown in Fig.~1(a), which can be numerically obtained using the GME method. As shown in Fig.~1(b), at frequency $\nu_1$, the group velocities for different wavevectors are nearly aligned along the $\Gamma\textrm{M}$ direction. In contrast, at frequency $\nu_2$ where the iso-frequency contour passes through the X-point, the group velocities display a broader direction distribution. For comparison, at frequency $\nu_3$, the IFC is nearly circular, and the corresponding group velocities are almost isotropic.

\begin{figure*}
	\includegraphics[width=0.95\textwidth]{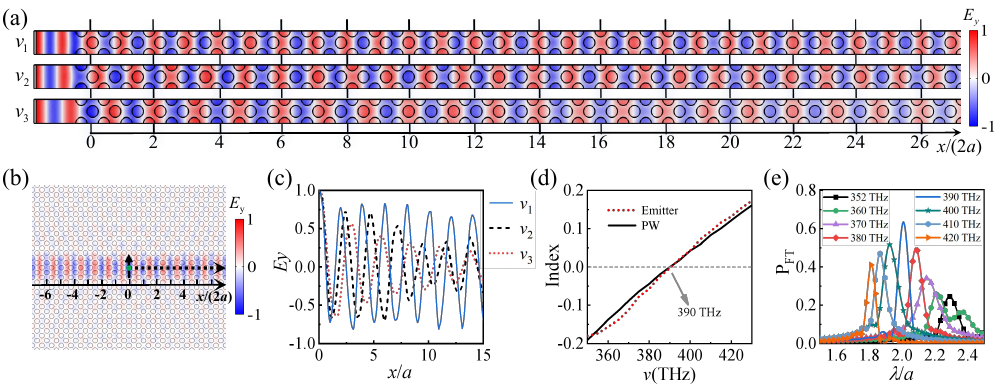}
	\caption{(Color online) \textbf{Unidirectional and low-loss emission.} (a) Variation of the electric field $E_y$ with distance for plane waves of different frequencies ($\nu_1\sim\nu_3$) incident from the left. (b) The electric field $E_y$ distribution generated by a dipole (green dot) with a frequency of $\nu_1$ on the central cross-section of the slab. (c) The normalized electric field $E_y$ excited by dipoles at different frequencies varies with distance along the black dashed line in (b). (d) The effective index of the electric fields excited by dipoles (red dots) and plane waves (black solid lines) are obtained from the phase variations of electric fields (a) and (b), respectively. (e) Fourier transform of normalized $E_y$ excited by a dipole at different frequencies. Horizontal axis: ratio of electric-field wavelength to lattice constant. Vertical axis: magnitude of the Fourier transform.}
\end{figure*}

These observations demonstrate that the group velocity distribution along an IFC provides an intuitive way to assess its straightness. However, to more precisely quantify the straightness of IFCs at different frequencies and avoid the cumbersome examination of group velocities at multiple wavevectors, we introduce a new metric called the straightness parameter $\Delta$, defined as
\begin{equation}
    \Delta = {d_{\rm max}}/{L},
\end{equation}  
where $L$ represents the arc length of the IFC in momentum space ($k_x, k_y \in \left[0, {\pi}/{a}\right]$), and $d_{\rm max}$ denotes the maximally perpendicular distance from each point on the IFC to the tangent line at the contour's midpoint. The parameter $\Delta$ serves as a dimensionless figure of merit, where $\Delta\to 0$ indicates perfect linearity (ideal unidirectivity), while larger values correspond to increased curvature. As shown in Fig.~1(c), this metric reveals distinctive behaviors near X-point: The unidirectional frequency $\nu_1$ exhibits minimal $\Delta$ values as denoted by a red dot ($\Delta\approx$ 0.02), confirming its exceptionally straight IFC geometry. In contrast, the saddle point and reference point show larger $\Delta$ represented by green and purple dots ($\Delta \approx$ 0.12 and 0.17), reflecting the characteristic curvature near the X-point. This quantitative analysis corroborates the qualitative observations from group velocity distributions while providing rigorous comparative metrics across the frequency spectrum.

To directly examine the correlation between emission directionality and IFC straightness ($\Delta$ parameter), we present in Fig.~1(d) the radiation patterns of an excited quantum emitter at three typical frequencies ($\nu_1\sim\nu_3$) in the center of a $29\times 29$ lattice. Each panel in Fig.~1(d) shows the normalized electric field intensity distribution in the central cross-section (excluding the divergent near-field region around the dipole), with a white arrow indicating the orientation of the excited dipole moment and color representing the field intensity. At the unidirectional frequency $\nu_1$ ($\Delta\approx$ 0.02), the radiation exhibits strong directionality along the $x$ direction, consistent with its quasi-linear IFC. As the frequency decreases to $\nu_3$, the emission pattern becomes increasingly isotropic as the IFC evolves toward a circular shape. This trend is quantitatively verified in Fig.~1(e) through polar plots of the electric field intensity measured at a fixed distance of $8\sqrt{2}a$ from the dipole, where $\nu_1$ demonstrates a highly directional emission pattern while the other two patterns show multidirectional emission.

\textbf{Unidirectional and low-less emission}. 
Analysis of the field patterns in Fig.~1(d) reveals distinct propagation characteristics: at the unidirectional frequency $\nu_1$, the electric field demonstrates remarkably slow and periodic decay along the $x$-direction, while at other frequencies ($\nu_2$ and $\nu_3$), the field decays rapidly to zero. This behavior resembles the unique properties of near-zero-index materials, where emitter radiation maintains nearly constant amplitude during propagation \cite{Minkov18prl}.

To further investigate this phenomenon, we examine plane wave excitation at three characteristic frequencies ($\nu_1\sim\nu_3$) by illuminating the slab structure from the left (same configuration as Fig.~1). Three panels in Fig.~2(a) show the resulting $y$-component electric field ($E_y$) distributions under three frequencies, respectively. Notably, at frequency $\nu_1$, the field exhibits stable $2a$-periodic oscillations with minimal amplitude decay along $x$-direction, while at other frequencies, the oscillations show different periods and significant attenuation. These observations suggest that the unidirectional frequency not only enables directional emission but may also support near-zero refractive index propagation with low loss in specific directions.

Complementing the plane wave analysis, Fig.~2(b) presents the $E_y$ distribution for a $y$-polarized dipole emitter with transition frequency $\nu_1$, positioned at the slab center. The radiation pattern shows striking similarity to the plane wave case in the $x$-direction [first panel of Fig.~2(a)], while exhibiting strong confinement in the $y$-direction due to the unidirectional emission characteristics discussed in the last subsection. Quantitative comparison in Fig.~2(c) demonstrates that only $\nu_1$ maintains a nearly constant field amplitude along $x$-direction, mirroring the plane wave behavior observed in Fig.~2(a).

To quantify these effects, we first introduce an effective refractive index $n_e^x$ for $x$-direction electric field propagation of both plane wave and emitter excitation, defined as
\begin{equation} 
E_y(x_0+m2a)=E_y(x_0)e^{in_e^xk_0m2a},
\end{equation} 
where $k_0 = \omega/c$ and $m$ takes integer values. As shown in Fig.~2(d), both emitter and plane wave excitations exhibit near-zero refractive index characteristics around $\nu_1$, explaining the observed $2a$-periodic phase preservation during propagation at this frequency.

While the effective refractive index $n_e^x$ provides insight into phase loss characteristics, it fails to capture two crucial aspects: radiation intensity loss at periodic positions and frequency-dependent periodicity variations. To address these limitations, we employ Fourier analysis of the normalized electric field $E_y$ (detailed in METHODS). Figure~2(e) presents the Fourier transform results under different frequencies, plotting transform amplitude against the wavelength-to-lattice-constant ratio ($\lambda/a$). The unidirectional frequency $\nu_1$ (blue curve) exhibits the strongest Fourier amplitude, confirming minimal propagation loss, while other frequencies show progressively greater losses. Remarkably, $\nu_1$ maintains a consistent $2a$-periodicity evidenced by its Fourier peak at $\lambda = 2a$, with frequency increases shifting this peak leftward (shorter wavelengths) and decreases shifting it rightward (longer wavelengths). This demonstrates that the $\nu_1$ spatial mode uniquely combines three advantageous properties: directional emission, robust $2a$-periodicity, and low propagation loss.

Fourier transform is a well-established analytical technique that enables comprehensive characterization of radiation properties through spatial electric field measurements along a unidirectional direction. By transforming the field distribution into the frequency domain and combining with the defined directivity parameter $\Delta$, this approach provides a unified framework for simultaneously quantifying directionality, energy loss, and oscillation periodicity. Compared to conventional emission directionality measurement methods, this Fourier-based technique offers superior precision in detecting subtle spectral features correlated with unidirectional emission while significantly simplifying experimental procedures by requiring only single-axis field measurements rather than full-space radiation pattern mapping. 

\textbf{Enhancing the single-emitter emission efficiency}. 
While the emitter with transition frequency $\nu_1$ exhibits excellent directionality and low propagation loss, its spectral detuning from the X-point frequency results in a relatively low spontaneous emission rate. This emission efficiency is quantified by the Purcell factor (PF)---the enhancement or suppression factor of the spontaneous emission rate for emitters in microstructures relative to free space. In our study, we specifically examine the PF for a $y$-polarized emitter, defined as
\begin{equation}
	\begin{split}
     \mathrm{PF} =\frac{\Gamma}{\Gamma_0} =\frac{6\pi c}{\omega_0}{\rm Im}[ {G}_{yy}(\vec r,\vec r,\omega)],
    \end{split}
\end{equation}
where the Green's function ${G}_{yy}$ is related to the electric field of the dipole at its own position $\vec r$, and a detailed calculation of the spontaneous emission of the dipole emitter is provided in METHODS.

In our previously studied PhC slab, saddle points in the band structure significantly enhance the PF through their high density of states, though this comes at the expense of reduced emission directionality as established earlier. This fundamental trade-off is clearly demonstrated in Fig.~3(a), where the PF for a dipole located at the center of the central air hole (blue solid line) shows stronger enhancement near the saddle-point frequency (shaded area) compared to the optimally directional frequency $\nu_1$ (indicated by black dashed line). The red dotted line represents the PF when the dipole is located
at the symmetric center of the dielectric slab, and although the PF is larger than that when the dipole is placed in the air hole, it causes the destruction of directionality, with a detailed description provided in the DISCUSSION section. To overcome this limitation, we strategically tuned the slab’s dielectric constant ($\varepsilon_r$) to simultaneously improve both emission directionality and efficiency. As shown in Fig.~3(b), increasing $\varepsilon_r$ systematically enhances the straightness of IFCs at saddle points, with the red data points tracking this evolution toward near-ideal linearity. Complementary Fourier analysis in Fig.~3(c) reveals that higher $\varepsilon_r$ values (comparing $\varepsilon_r=4$, blue solid line to $\varepsilon_r=32$, cyan dotted line) yield not only sharper directional peaks but also reduced propagation loss through increased Fourier amplitudes. The combined improvements are quantitatively summarized in Fig.~3(d), where optimized dielectric tuning achieves enhanced PF while maintaining excellent directionality effectively transforming saddle-point frequencies into optimal operational points that combine the benefits of high DOS regions with the directional emission characteristics of carefully engineered photonic band structures.
\begin{figure}
	\includegraphics[width=0.5\textwidth]{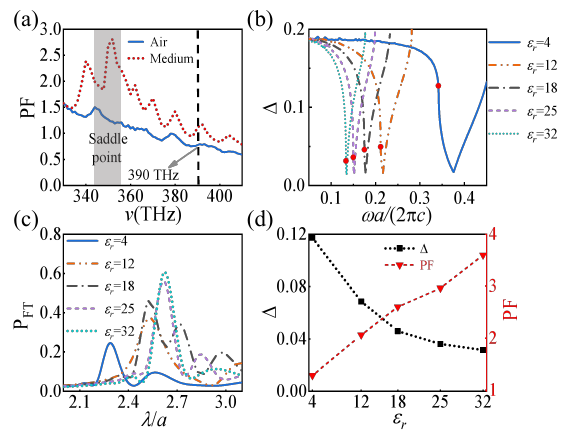}
	\caption{(Color online) \textbf{Enhancing the Purcell factor}. (a) The PF when dipoles of different frequencies are located at the center of the hole (blue solid line) and the symmetric center of the medium (red dotted line) in a slab with a dielectric constant of $\varepsilon_r=4$. The shaded part represents the frequencies near the saddle-point frequency. (b) The straightness of IFC at various frequencies for the first energy band of TE-like modes in PhC slabs with different dielectric constants. The red dots on each curve correspond to the straightness of IFC at the saddle-point frequency. (c) The Fourier transforms of the field $E_y$ excited by dipoles at the saddle-point frequency for PhC slabs with different dielectric constants. (d) As the dielectric constant increases, the straightness of the IFC at the saddle-point frequency (black dashed line with squares) gradually decreases, while the PF of the dipole at the saddle-point frequency (red dashed line with inverted triangles) gradually increases.}
\end{figure}

\begin{figure}
	\includegraphics[width=0.5\textwidth]{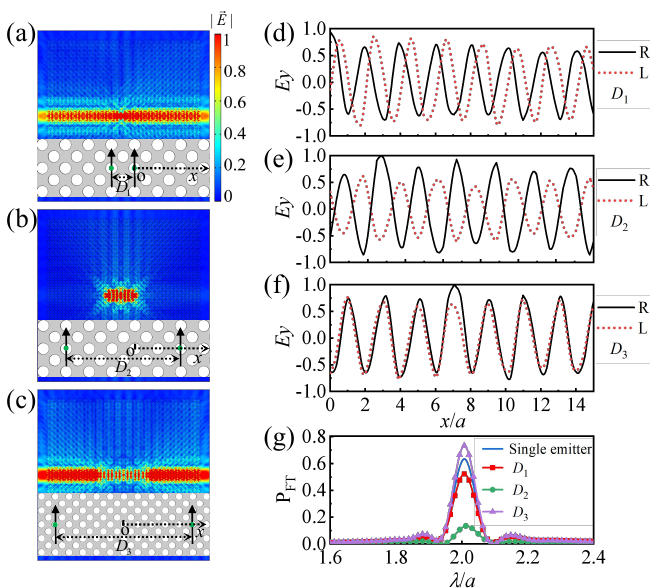}
	\caption{(Color online) \textbf{Interference effects of two emitters}. (a), (c), and (e) depict the electric field intensity excited by two dipoles positioned at different locations on the central cross-section of the slab, as indicated in the inset. Both dipoles are located at the centers of the air holes, with distances $D_1=\sqrt2a$, $D_2=5\sqrt2a$, and $D_3=10\sqrt2a$. (b), (d), and (f) show the variation of the electric field $E_y$ individually excited by two dipoles along the dashed line $o$-$x$ direction in the inset with distance, where the red dotted line corresponds to the left dipole and the blue solid line represents the right dipole. (g) The comparison between the amplitude of the one-sided spectral Fourier transform of the electric field $E_y$ excited by two dipoles with different spacings and that of a single dipole (blue solid line). The frequency of the dipoles is $\nu=390~\rm THz$.}
\end{figure}


\textbf{Collective effects of multiple emitters}. While our previous investigations have primarily focused on single-emitter configurations within the photonic crystal slab system, extending this study to multiple emitters reveals both new physical phenomena and engineering challenges. The study of multiple emitter systems is of fundamental importance in nanophotonics, as collective interactions and near-field interference can lead to remarkable phenomena, including superradiance and subradiance. In our PhC slab platform, although the straightness of IFCs and Fourier-transform-amplitude analysis effectively characterize single-emitter emission properties, the collective behavior of multiple emitters introduces rich many-body effects that significantly modify radiation characteristics. Specifically, the constructive/destructive interference of unidirectional fields and the modified collective emission rates jointly govern the resulting emission patterns, which will be the focus of this section. Through investigation of these two effects, we will demonstrate how emitter arrangements can be engineered to achieve tailored directional emission with enhanced radiation characteristics.

We first investigate the interference effects of two emitters with different spatial configurations along the $x$-direction, focusing exclusively on the optimal unidirectional frequency $\nu_1$ where all emitters maintain $y$-polarization and are precisely positioned at air hole centers (the impact of off-center positioning is discussed in DISCUSSION). Figures~4(a)-(c) display the total radiation patterns for three distinct emitter separations ($D_1=\sqrt{2}a$, $D_2=5\sqrt{2}a$, and $D_3=10\sqrt{2}a$), while Figs.~4(d)-(f) show the corresponding $x$-axis field distributions ($E_y$) from the right ($R$, black solid lines) and left ($L$, red dotted lines) emitters. For the nearest-neighbor case ($D_1=\sqrt{2}a$), the interference pattern deviates slightly from the characteristic $2a$ periodicity of single-emitter radiation, resulting in only $1.5\times$ intensity enhancement. As $D_2\approx 7a$ (approximately an odd multiple of the half-period of the single-emitter case), the out-of-phase interference between emitters leads to complete field cancellation, as clearly demonstrated in Figs.~4(b) and 4(e). Remarkably, the $D_3\approx 14a$ configuration (approximately an even multiple of $2a$) achieves nearly perfect phase synchronization [Fig.~4(f)], yielding a $2\times$ field amplitude enhancement compared to a single emitter. Fourier analysis in Fig.~4(g) further confirms that the $D_3$ configuration not only enhances emission intensity but also improves directionality beyond single-emitter performance. These findings demonstrate that strategically arranged emitter arrays in our PhC slab can simultaneously achieve superior directionality and scalable intensity enhancement proportional to the number of emitters.

\begin{table}[htbp]
    \centering
    \caption{The second-order correlation function $g^{(2)}(0)$ is shown for a chain of six emitters as a function of emitter separation (specifically, $D_1$, $D_2$, and $D_3$) under three configurations: (i) centered in air holes, (ii) at the symmetry center of the dielectric medium, and (iii) in vacuum as a reference.}
    \label{tab:spacing}
    \renewcommand{\arraystretch}{1.5}
    \label{tab:fixed_width}
    \begin{tabular}{>{\centering\arraybackslash}p{2cm}
    >{\centering\arraybackslash}p{2cm}
    >{\centering\arraybackslash}p{2cm}
    >{\centering\arraybackslash}p{2cm}}
        \hline
        & $D_1$ & $D_2$ & $D_3$ \\
        \hline
        (i)   & 1.010   &  1.179  & 1.104 \\
        (ii)  & 0.935   & 0.938   & 0.911 \\
        (iii) & 0.855   &  0.835  & 0.833 \\
        \hline
    \end{tabular}
\end{table}

The interference patterns observed in the radiation fields of two emitters (as described above) represent a classical effect, which characterizes the spatial interference of multi-emitter systems. Beyond this spatial phenomenon, another critical collective effect emerges: the modified emission rates, reflecting the temporal dynamics of cooperative radiation. Superradiance occurs when the collective radiation intensity of multiple emitters exceeds the sum of their individual emissions---a signature of enhanced quantum coherence. This effect typically arises in emitter arrays with long-range dissipative interactions.

The minimal condition for superradiance is quantified by $g^{(2)}(0)>1$, with the second-order correlation function being \cite{Ren24prapp}
\begin{equation}
	\begin{split}
     g^{(2)}(0) = 1 + \frac{\operatorname{Tr}\left( \boldsymbol{\Gamma}^2 \right) - 2 \operatorname{Tr}\left( \boldsymbol{\Gamma}_\text{d}^2 \right)}{\left( \operatorname{Tr}\boldsymbol{\Gamma} \right)^2},
    \end{split}
\end{equation}
where $\boldsymbol{\Gamma}=(\gamma_{ij})$ is the incoherent coupling matrix with $\gamma_{ij}$ being dissipative coupling between the $i$th and $j$th emitters, and $\boldsymbol{\Gamma}_\text{d} \equiv \operatorname{diag}(\gamma_{11}, \gamma_{22}, \ldots, \gamma_{NN})$ is the diagonal matrix of $\boldsymbol{\Gamma}$.

To systematically investigate collective emissions in multi-emitter systems, we calculated the second-order correlation function $g^{(2)}(0)$
for a chain of six emitters positioned at high-symmetry points of the square lattice with different separation distances [Figs.~4(a)-(c)] under three configurations: (i) centered in air holes, (ii) at the symmetry center of the dielectric medium, and (iii) placed in vacuum as a reference (see Table 1 for quantitative results). As summarized in Table 1, configuration (i) exhibited $g^{(2)}(0)>1$ for all tested separations ($D_1$, $D_2$, $D_3$), confirming superradiance, while this enhancement weakened for configuration (ii) and disappeared completely in vacuum (iii) where all $g^{(2)}(0)$ values remained below 1. These results demonstrate that at optimal frequency $\nu_1$, the PhC slab platform simultaneously achieves unidirectional emission with low loss, scalable intensity enhancement, and superradiant emission, highlighting the unique synergy between photonic band engineering and multi-emitter coupling that enables unprecedented control over both spatial and temporal coherence in quantum emitter arrays.


\begin{figure*}
	\includegraphics[width=0.85\textwidth]{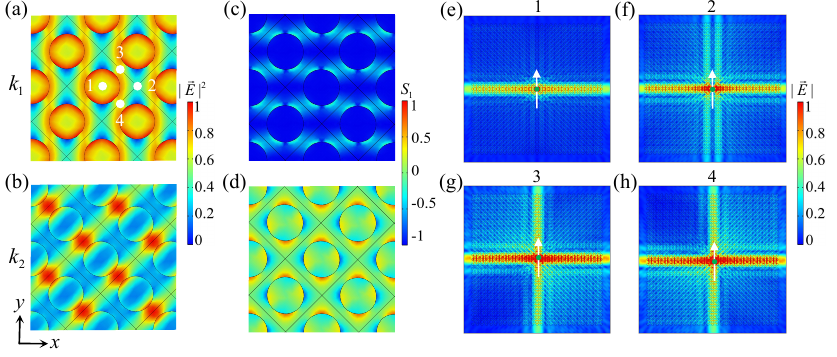}
	\caption{(Color online) \textbf{Directional emission and the interaction between field polarization and emitter position.} (a) and (b) correspond to the squared norm of the electric field ($V^2/m^2$) of the two eigenmodes corresponding to the unidirectional frequency $\nu_1$ in Fig.~1(a), with their wave vectors being $k_1$ and $k_2$, respectively. The four white dots in (a) denote four special positions: 1 represents the center of the air hole, 2 represents the symmetry center of the dielectric medium, and 3 and 4 represent the midpoints of the dielectric slabs between the nearest neighboring holes. (c) and (d) represent the Stokes parameter $S_1$ corresponding to the eigenmodes with wave vectors $k_1$ and $k_2$, respectively.  (e), (f), (g), and (h) respectively show the electric field intensity excited by a single dipole located at these four special positions. All dipoles are polarized along the $y$-direction, as indicated by the white arrows in the figures.}
\end{figure*}

\section{discussion}
\textbf{Scalable multi-emitter emission: symmetry requirements and robust directivity}. 
To complement our previous investigation of center-positioned emitters in the RESULTS section, we extend the discussion to off-center configurations by calculating the coupling strengths between the emitter and the two eigenmodes ($k_1$ and $k_2$) at the unidirectional frequency $\nu_1$ for different positions: 1 denotes the center of the air hole, 2 denotes the symmetry center of the medium, and 3 and 4 denote the midpoints of the dielectric slabs between the nearest neighboring holes [Figs.~5(a) and~5(b), respectively]. It can be seen from these four special positions that positions 1 and 2 possess $C_{4v}$ symmetry, while positions 3 and 4 have $C_{2v}$ symmetry. Therefore, by combining the squared norm of the electric field at these four positions in Figs.~5(a) and~5(b), it can be concluded that the emitters at positions 1 and 3 have strong coupling with the mode corresponding to $k_1$, but weak coupling with the mode corresponding to $k_2$, while positions 2 and 4 correspond to mixed modes of $k_1$ and $k_2$. The mode corresponding to $k_1$ is a unidirectional mode along the $\Gamma\textrm{M}$ direction, while $k_2$ is a mode deviating from the $\Gamma\textrm{M}$ direction. Therefore, emitters at positions 1 and 3 at the unidirectional frequency $\nu_1$ mainly excite the unidirectional mode, whereas emitters at other positions will have coupling strengths comparable to those of the $k_1$ and $k_2$ modes, thus destroying the directionality of the radiation pattern. In addition to the influence of the emitter position on the directionality of radiation, the polarization direction of the emitter is equally important, so we describe it using the Stokes parameter $S_1$, which is defined as \cite{Barón24np}
\begin{equation}    
S_1 = \frac{E_x^2 - E_y^2}{|\vec{E}|^2},
\end{equation}
where $E_x$ and $E_y$ are the components of the eigenfield along the coordinate axes in Fig.~5, respectively; when $S_1$ is 1 (-1), it indicates that the eigenfield at this position is mainly polarized in the $x$ ($y$) direction, and dipoles polarized in the $x$ ($y$) direction will have strong coupling with it.
Figs.~5(c) and~5(d) show the $S_1$ parameters of the modes corresponding to the wave vectors $k_1$ and $k_2$, respectively. It can be seen from the figures that at the four special positions, the $k_1$ mode is dominated by the $E_y$, while the $k_2$ mode has comparable proportions of $E_x$ and $E_y$. However, only the $S_1$ parameter at position 1 is closest to the extreme value, meaning that the $y$-polarized dipole at this position has strong coupling with the $k_1$ mode and radiates only a single channel, exhibiting good directionality; whereas for position 3, although it also has strong coupling only with the $k_1$ mode, its $S_1$ parameter is not an extreme value, causing the $y$-polarized dipole at this position to excite two radiation channels and thus destroying the directionality. As shown in Figs.~5(e), 5(f), 5(g), and 5(h) show the electric field intensity excited by the dipoles polarized in the $y$ direction placed at the four special positions, respectively, with the white arrows in the figures indicating the polarization directions of the dipoles. 
As can be seen in the figure, as expected, for dipoles polarized in the $y$-direction, the radiation field maintains single-channel directionality only when they are located at position 1. For the other three positions, the radiation channel along the $y$-direction generates a weaker field intensity, and the directionality of the radiation is destroyed. However, since the emitters are located in the medium with a larger PF, the radiation field intensity is higher. In summary, placing the emitter in the air hole is the optimal choice for unidirectional emission.

\begin{figure}
	\includegraphics[width=0.5\textwidth]{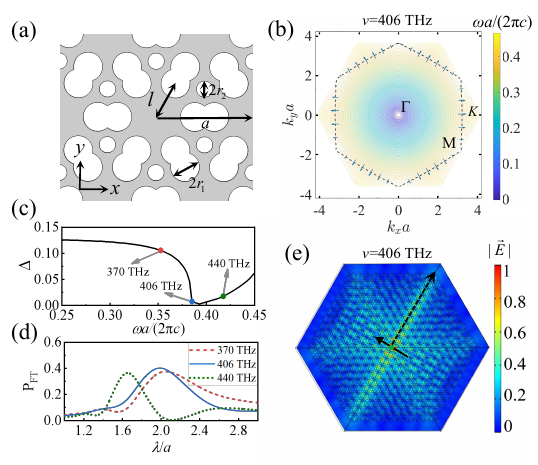}
	\caption{(Color online) \textbf{Directional spontaneous emission in triangular lattice}. (a) is a schematic of a triangular lattice PhC slab with lattice constant $a=280.44~\rm nm$, $l=0.4a$, $r_1=0.15a$, $r_2=0.0833a$, and thickness $d=0.25a$. The shaded regions have a dielectric constant of $\varepsilon_\textrm{Gap}=10.5625$, and the white holes are air holes. (b) shows the electric field intensity (the normalized electric field intensity after removing the field near the dipole) excited by dipoles with a polarization direction at an angle of $5\pi/6$ from the horizontal direction, corresponding to the frequency of $406~\rm THz$ at point $\textrm{M}$. (c) show the constant frequency contours of the first band for the TE-like, where the black dashed line corresponds to the constant frequency contour at the M-point frequency, and the blue arrows indicate the group velocity. (d) shows the straightness calculated for the IFC at each frequency in (c), where the blue dots represent the straightness at the saddle-point frequency. (e) The one-sided spectral Fourier transform of the normalized electric field component in the polarization direction along the black dashed line in (b), with the blue solid line representing $406~\rm THz$ (the saddle-point frequency).}
\end{figure}
\textbf{Directional emission in a triangular lattice}. Our previous discussions focused on square lattices, but existing studies confirm that unidirectional radiation can also be achieved in triangular-lattice PhC slabs \cite{Barón24np}. As illustrated in Fig.~6(a), a proposed triangular lattice structure features a unit cell with two air holes of radii $r_1$ (large) and $r_2$ (small), offset by the distance $l$ from the center of the unit cell. The IFC analysis for the first TE-like band [Fig. 6(b)] reveals that at the saddle-point frequency ($\nu=406$ THz), the IFC passing through the high-symmetry M-point exhibits exceptional straightness, with nearly aligned group velocity vectors. 

The straightness of the M-point IFC in Fig.~6(c), quantified by the straightness parameter $\Delta$, reaches its minimum at $\nu = 406$ THz (saddle-point frequency), suggesting optimal directional emission at this frequency. However, Fourier analysis of the radiation field [Fig.~6(d)] reveals that while the amplitude at $\Delta$-minimizing frequency is the highest among all frequencies, it remains inferior to the optimal case in square lattices studied previously, indicating suboptimal practical directivity. This discrepancy is further evidenced by the dipole radiation pattern in Fig.~6(e): the emitter at the saddle-point frequency generates a field that decays rapidly along the corresponding direction of $\Gamma$K  in real space, aligning with the IFC-predicted directionality but with weaker unidirectionality than its square-lattice counterparts.

These results confirm that triangular lattices can indeed support engineered unidirectional emission, albeit with distinct band-structure mechanisms. The reduced directivity compared to square lattices stems from fundamental geometric differences: in square lattices, orthogonal $\Gamma$M directions enable independent channel control, whereas triangular lattices exhibit nonorthogonal $\Gamma$K directions [Fig. 6(b)], causing mutual interference between emission channels.

\section{methods}
\textbf{Guided-mode expansion method}. The GME method is a theoretical approach for investigating the light propagation characteristics in PhC slabs, and it is mainly used to calculate photonic band gaps, mode dispersion, and diffraction losses \cite{Andreani06prb,Minkov20acsp,Zanotti24cpc}. The eigenmodes of the uniform slab are divided into two categories: guided modes (with fields confined in the core layer and propagating within it) and radiative modes (with fields leaking into the cladding layers accompanied by energy radiation). The core of the GME method lies in the use of the guided modes of the uniform slab as basis functions: When the periodic structure weakly modulates the dielectric constant of the uniform slab, the actual electromagnetic modes of the periodic slab can be approximately expressed as a linear superposition of the guided modes of the uniform slab. By solving the guided modes of such an approximately uniform dielectric constant slab, substituting them into Maxwell's equations, and combining with mode orthogonality, the eigenmodes of the periodic photonic crystal slab can be solved.
For the photonic crystal slab studied in this paper, which has a dielectric plate (with a thickness of $d$) as the core layer and air as the upper and lower cladding layers, the average dielectric constants of the corresponding uniform slab are $\bar{\varepsilon}_1$ (upper and lower cladding layers) and $\bar{\varepsilon}_2$ (core layer), respectively. When the magnetic field has a harmonic time dependence $\mathbf{H}(\mathbf{r}, t)=\mathbf{H}(\mathbf{r})e^{-i\omega t}$, Maxwell's equations can be transformed into a second-order equation containing only the magnetic field,
\begin{equation}    
\nabla \times \left[ \frac{1}{\varepsilon(\mathbf{r})} \nabla \times \mathbf{H}(\mathbf{r}) \right] = \frac{\omega^2}{c^2} \mathbf{H}(\mathbf{r}),
\end{equation}
 with the condition that $\nabla \cdot \mathbf{H} = 0$. Expand the magnetic field in a set of orthonormal bases as
\begin{equation}  
\mathbf{H}(\mathbf{r}) = \sum_{\mu} c_{\mu} \mathbf{H}_{\mu}(\mathbf{r}),
\end{equation}
$\{\mathbf{H}_{\mu}(\mathbf{r})\}$ represents a set of guided modes in a homogeneous dielectric slab, which satisfy the orthonormalization condition
\begin{equation}  
\int \mathbf{H}_{\mu}^{*}(\mathbf{r}) \cdot \mathbf{H}_{\nu}(\mathbf{r}) \, d\mathbf{r} = \delta_{\mu\nu}.
\end{equation}
The guided modes of the uniform slab can be separated into two orthogonal families: transverse electric (TE) and transverse magnetic (TM) modes. The eigenfrequencies of the guided modes in these two families can be obtained by the following implicit equations:
\begin{equation}  
\begin{split}
2q_{\mu}\chi_{\mu}\cos(q_{\mu} d)
+(\chi_{\mu}^{2}-q_{\mu}^2) \sin(q_{\mu} d) = 0
\end{split}
\end{equation}
for transverse electric (TE) polarization, and
\begin{equation}  
\begin{split}
2\frac{q_{\mu}\chi_{\mu}}{\bar{\varepsilon}_2\bar{\varepsilon}_1}\cos(q_{\mu} d) 
+\left( \frac{\chi_{\mu}^{2}}{\bar{\varepsilon}_1^2} - \frac{q_{\mu}^2}{\bar{\varepsilon}_2^2} \right) \sin(q_{\mu} d) = 0
\end{split}
\end{equation}
for transverse magnetic (TM) polarization, where
\begin{equation}
\chi_{\mu}= \left( \text{g}^2 - \bar{\varepsilon}_1 \frac{\omega_\mu^2}{c^2} \right)^{1/2}
\end{equation}
and
\begin{equation}
q_{\mu}= \left( \bar{\varepsilon}_2 \frac{\omega_\mu^2}{c^2} - \text{g}^2 \right)^{1/2}
\end{equation}
represent the imaginary parts of the wave vectors in the upper and lower cladding layers and the real part of the wave vector in the core layer, respectively. Guided modes are labeled by the wave vector $\mathbf{g}$ ($\mathbf{g}=\text{g}\hat{\text{g}}$ is a two-dimensional wave vector in the $xy$-plane, where $\text{g}$ is the modulus of the vector and $\hat{\text{g}}$ is the unit vector) and mode index $\alpha$, which are combined into a single index $\mu=(\mathbf{g},\alpha)$, with $\omega_\mu$ being the frequency of the guided mode. 

Substituting Eq.~(8) into Eq.~(7) can transform Eq.~(7) into a linear eigenvalue problem. 
\begin{equation}  
\sum_{\nu} \mathcal{H}_{\mu\nu} c_{\nu} = \frac{\omega^2}{c^2} c_{\mu},
\end{equation}
by solving this eigenvalue equation, the eigenvalues of the photonic crystal slab and their corresponding eigenmodes can be obtained, where the matrix $\mathcal{H}_{\mu\nu}$ is given by
\begin{equation}  
\mathcal{H}_{\mu\nu} = \int \frac{1}{\epsilon(\mathbf{r})} \left( \nabla \times \mathbf{H}_{\mu}^{*}(\mathbf{r}) \right) \cdot \left( \nabla \times \mathbf{H}_{\nu}(\mathbf{r}) \right) d\mathbf{r}.
\end{equation}
Once the magnetic field is solved, the electric field can be obtained using the following formula:
\begin{equation} 
\mathbf{E}(\mathbf{r}) = \frac{ic}{\omega \varepsilon(\mathbf{r})} \boldsymbol{\nabla} \times \mathbf{H}(\mathbf{r}).
\end{equation}

\textbf{Classical Green’s function and quantum spontaneous emission}. Using the Green function can facilitate the solution of electromagnetic field problems with sources, and through Maxwell’s equations, one can obtain \cite{Zhu23oe,Ren24prapp,Chu25npj}
\begin{equation}  
\nabla \times \nabla \times \overset{\leftrightarrow}{G}(\vec r_1,\vec r_2,\omega) - \frac{\omega^2}{c^2} \varepsilon(\vec r_1, \omega) \overset{\leftrightarrow}{G}(\vec r_1,\vec r_2,\omega) = \mathbf{I} \delta(\vec r_1 - \vec r_2),
\end{equation}
the Green tensor $\overset{\leftrightarrow}{G}(\vec r_1,\vec r_2,\omega)$ represents the electric field at position $\vec r_1$ due to a source located at $\vec r_2$ with frequency $\omega$, and the distance between them is $r=|\vec r_1-\vec r_2|$. In this work, the dipole moments in the square lattice are all directed along the $y$ axis and have the same electric dipole moment. Therefore, the Green's function can be obtained by solving for the electric field:
\begin{equation}
	\begin{split}
     \vec\mu_1^\ast\cdot \overset{\leftrightarrow}{G}(\vec r_1,\vec r_2,\omega)\cdot\vec\mu_2=-E_y(\vec r_1)_{\vec r_2},
    \end{split}
\end{equation}
where $E_y(\vec r_1)_{\vec r_2}$ is the $y$ component of the electric field at $\vec r_1$ emitted from the emitter located at $\vec r_2$. Using the Green's function, it can be obtained that the spontaneous emission rate of the dipole at position $\vec r_1$ is 
\begin{equation}
	\begin{split}
     \Gamma=\frac{2\omega_0^2}{\varepsilon_0\hbar c^2}{\rm Im}[\vec\mu_1^\ast\cdot \overset{\leftrightarrow}{G}(\vec r_1,\vec r_1,\omega)\cdot\vec\mu_1].
    \end{split}
\end{equation}
As is well known, the spontaneous emission rate of a dipole in vacuum is
\begin{equation}
	\begin{split}
     \Gamma_0 = \frac{\omega_0^3 |\vec\mu_1|^2}{3\pi \varepsilon_0 \hbar c^3}.
    \end{split}
\end{equation}

\textbf{Fourier Transform}.
The Fourier Transform (FT) is a core tool in mathematics and physics, whose essence is to convert signals from the time domain or spatial domain to the frequency domain for studying the spectral structure and variation laws of signals \cite{Cooley65moc,Bracewell00book,Goodman05book}. In this paper, we utilize the FT to characterize the attenuation of the electric field with distance. Given the normalized electric field variation with distance $E_y(r)$, its discrete Fourier transform is denoted as
\begin{equation}
\mathcal{F}(k) =\int_{-\infty}^{\infty}E_y(r)e^{-ikr}dr,
\end{equation}
where $k$ is the wavenumber. To ensure energy conservation and make the amplitude spectrum physically meaningful, the frequency spectrum of negative frequencies needs to be discarded, and the amplitude of the positive frequency components should be doubled to obtain the unilateral spectrum
\begin{equation}
P_{FT}(k) = 
\begin{cases} 
2\vert\mathcal{F}(k)\vert & k > 0 \\
\vert\mathcal{F}(k)\vert & k = 0 .
\end{cases}
\end{equation}
As shown in the RESULTS, when discussing problems involving multiple dipoles, it is generally required to obtain the periodic variations of the dipole radiation field. Therefore, the wavelength-domain Fourier transform is more convenient

\begin{equation}
P_{FT}(\lambda) = 2\vert\int_{-\infty}^{\infty} E_y(r)e^{-i \frac{2\pi}{\lambda}r}dr\vert\quad(\lambda\neq\infty),
\end{equation}
where $\lambda$ is the wavelength of the electromagnetic wave propagating in space.


\begin{acknowledgments}
	This work was supported by NSF-China (Grant Nos. 11904078 and 11575051), GRF (Grant Nos. 17310622 and 17303023) of Hong Kong, Hebei NSF (Grant Nos. A2021205020 and A2019205266), and Hebei 333 Talent Project (B20231005). JR was also funded by the project of the China Postdoctoral Science Foundation (Grant No. 2020M670683).
\end{acknowledgments}




\end{document}